\documentclass{aa}
\usepackage{graphicx}
\input{psfig.sty}
\begin{document}
   \title{Surface photometry and structure of high redshift disk
        galaxies in the HDF-S NICMOS field}
   \author{A. Tamm \and
           P. Tenjes}
   \offprints{P. Tenjes}
   \institute{Institute of Theoretical Physics, Tartu University,
              T\"ahe 4, Tartu, 51010 Estonia;\\
              Tartu Observatory, T\~oravere, Tartumaa,
             61602 Estonia\\
              \email{atamm@ut.ee; ptenjes@ut.ee}
              }
   \date{Received ..... ; accepted .............}

   \abstract{

   We report on a photometric study of a sample of 22 disk galaxies in
the Hubble Deep Field South NICMOS parallel field. The redshift
range of the galaxies is $z=0.5-2.6$. We use deep NICMOS J and H
band and STIS open mode images, taken as part of the HDF-S project,
to construct rest-frame $B$-profiles and $(U-V)$ color profiles of
the galaxies. Before fitting isophotes, images are deconvolved with
PSF. Derived surface brightness profiles are approximated by
S\'{e}rsic luminosity distribution. Significantly large population
of disks cannot be represented by an exponential disk, but this can
be well done by S\'{e}rsic law, if $n<1$. This might be the same
phenomenon which has earlier been referred to as truncation of
disks. Parameter $n$ does not vary significantly with redshift.
Galactic sizes decrease with redshift as $r_e(z)/r_e(0) = 1-0.26z$.
The rest frame $(U-V)$ color shows a clear decrease at $z\approx2$,
concordantly with the understanding of more intense star formation
at earlier epochs. Color gradients $\Delta(U-V)/\Delta r$ are small
and roughly constant at $z<2$. At $z>2$, dominantly positive
gradients appear, possibly indicating centrally concentrated
star-formation. On the basis of $(U-V)$ color and chemical evolution
models, the disks observed at $z\sim 2.5$ have formed between
$z=3.5-7$. Scale radii $r_e$ of the galaxies correlate with the
scale surface brightnesses $\mu_e$ for the sample. None of the
studied parameters shows clear dependence on absolute B luminosity
for the galaxies.

   \keywords{galaxies: photometry -- galaxies: fundamental parameters --
             galaxies: high redshift -- galaxies: spiral -- galaxies:
             structure}
             }
\authorrunning{Tamm \& Tenjes}
\titlerunning{Photometry of high-redshift disk galaxies}
   \maketitle
%

\section{Introduction}

Comparison of the structure of galactic disks at intermediate and
high redshifts with the disks resulting from N-body simulations
allows us to test the corresponding algorithms and to specify
physical processes involved in disk formation. General properties
of both elliptical and disk galaxies are rather well reproduced by
modeling (e.g. White \& Rees \cite{white}; Navarro \& White
\cite{nav}; Abadi et al. \cite{abadi}; Bell et al. \cite{bell};
Governato et al. \cite{gove}; Robertson et al. \cite{robe};
Nagamine et al. \cite{naga}). The initial shortcomings have been
corrected by now and it has become possible to model the formation
of extended, rotationally supported disks. On the other hand, good
models for calculating the formation and evolution of galaxies
involve a great number of free parameters (see e.g. Cole et al.
\cite{cole}).

To test and constrain the simulations carried out at increasing
resolutions, detailed observational studies of individual high
redshift galaxies are indispensable. In a number of papers,
structural parameters of galaxies have been statistically studied
at a wide range of redshifts. Luminosity function has been derived
out to $z\sim 1.5-2$ by Chen et al. (\cite{chen}) and Ilbert et
al. ({\cite{ilbe}); to $z\sim 3.5$ by Poli et al. (\cite{poli})
and Giallongo et al. (\cite{gial}); to $z\sim 5$ by Gabasch et al.
(\cite{gaba}). General morphologic studies and statistics out to
redshift $z\approx3$ and even further have been conducted by
Bunker et al. (\cite{bunk}), van den Bergh et al. (\cite{bergh}),
Corbin et al. (\cite{corb}), Bouwens et al. (\cite{bouw}), Cassata
et al. (\cite{cass}) and many others.

Due to hierarchical clustering, galactic sizes should grow with
time (e.g. Fall \& Efstathiou \cite{fall}). An analysis of
galactic size distribution allows us to constrain certain
parameters in galaxy evolution simulations (feedback etc., see
Cole et al. \cite{cole}). The observations do not clearly confirm
this prediction as yet. For example, galaxy size evolution with
redshift has been studied by Ferguson et al. (\cite{ferg}) and
Trujillo et al. (\cite{truj}, \cite{truj2}). They show that, at a
given luminosity, decrease of the sizes of galaxies with redshift
is clearly present. On the other hand, Ravindranath et al.
(\cite{ravi}) and Cassata et al. (\cite{cass}) do not detect
significant evolution. Serious sources of controversies are
selection effects and difficulties in conducting photometric
analysis of galaxies at high redshifts, both caused mainly by
cosmologic dimming of surface brightness by a factor of $(1+z)^4$.

While there is a growing inflow of data concerning general
statistical properties of high redshift galaxies, just a few
papers have concentrated on the internal properties of distant
disk galaxies, e.g. kinematics, detailed surface brightness and
color distribution. For example, rotation curves for distant disk
galaxies ($z \le 1$) have been measured by Vogt et al.
(\cite{vogt1}, \cite{vogt2}), Erb et al. (\cite{erb}) and B\"{o}hm
et al. (\cite{bohm}). Moth \& Elston (\cite{moth}) have studied
HDF-N field and constructed rest-frame $(UV_{218}-U_{300})$ color
profiles and rest-frame $B$ surface brightness profiles for 83
galaxies at $z=0.5-3.5$. They report color gradient
$\Delta(UV_{218}-U_{300})/\Delta r$ rising with redshift, which
suggests that star formation has shifted inwards during the
evolution of galaxies.

In Tamm \& Tenjes (\cite{TT1}, \cite{TT2}), we combined luminosity
distribution with rotation curves to construct self-consistent
mass distribution models for disk galaxies out to $z\approx1$. We
noted that for all the 7 galaxies studied, the usual exponential
disk assumption gave poor fit to the luminosity profiles; a more
steep cut-off at outer radii, provided by the S\'{e}rsic
(\cite{sersic}) index $n<1$, was needed. Was this because of
selection effects, evolution of disk parameters with time or
anything else? With this question in mind, in the present paper we
concentrate on the luminosity and color profiles of disk galaxies
at a redshift range as wide as possible, in an attempt to detect
evolutionary effects of several disk parameters with a particular
emphasis on the shape of the profiles. To achieve this, our main
task is to acquire surface brightness profiles as well as color
profiles for a sample of disk galaxies, introducing as few
selection effects as possible.

We have chosen a small field in the southern sky, the Hubble Deep
Field (HDF) South NICMOS parallel field for the study. In addition
to the information offered by the multi-color observations of this
field, the choice of the southern sky provides comparison with the
well-studied HDF-N region.

In the present work we take $H_0 =$ $\rm 65~km\ s^{-1}Mpc^{-1}$ and
$\Omega_0 = 1,$ $\Omega_m = 0.3,$ $\Omega_{\Lambda} = 0.7.$

\begin{table*}
   \caption[]{General information about the HDF-South NICMOS field observations}
   \label{tab1}
\begin{center}
\begin{tabular}{llllll}
\hline
Camera \& filter& Central        & Bandwidth & Total exposure& PSF width   & Projected pixel    \\
                & wavelength (nm)& (FWHM, nm)& time (s)      & (FWHM, $''$)& angular size ($''$)\\
\hline
STIS open mode  & 550            & 441       & 25900         & 0.065       & 0.025\\
NICMOS F110W    & 1132           & 588       & 108539        & 0.23        & 0.075\\
NICMOS F160W    & 1608           & 399       & 128441        & 0.23        & 0.075\\
\hline
\end{tabular}
\end{center}
\end{table*}

\begin{figure*}
\begin{center}
\includegraphics[width=16.5cm]{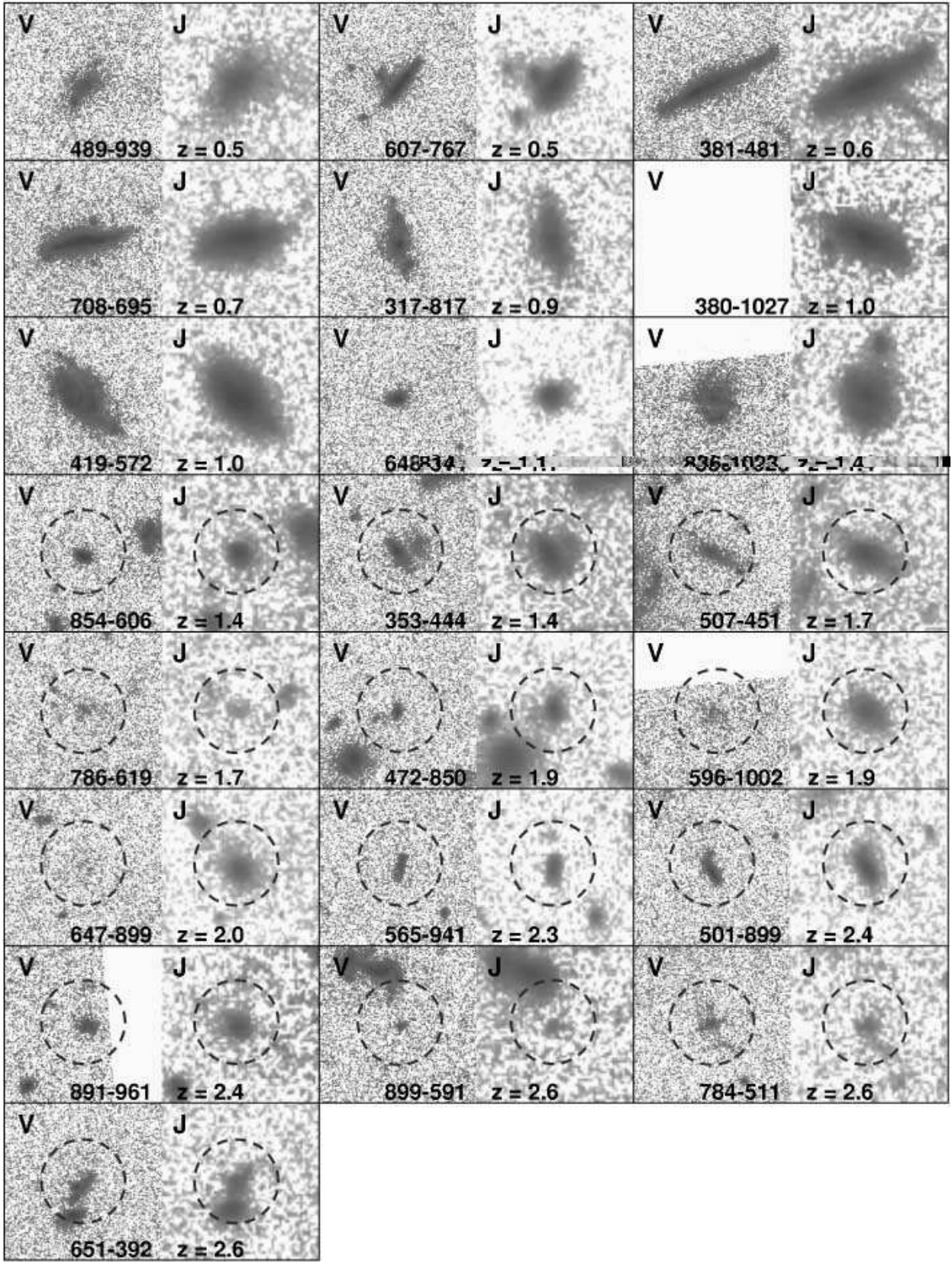}
\caption{Galaxies studied in the present paper as they appear on
STIS open mode ($\sim V$) and NICMOS $J$ passband images of the
HDF-S NICMOS field. For clarity, the studied galaxy has been
encircled on some images. The angular size of each image is
$4.5\times4.5$ arcseconds. $x$ and $y$ coordinates of NICMOS images
are used as galactic names; $z$ is photometric redshift.}
\label{fig1}
\end{center}
\end{figure*}

\begin{figure*}
\begin{center}
\includegraphics[width=17cm]{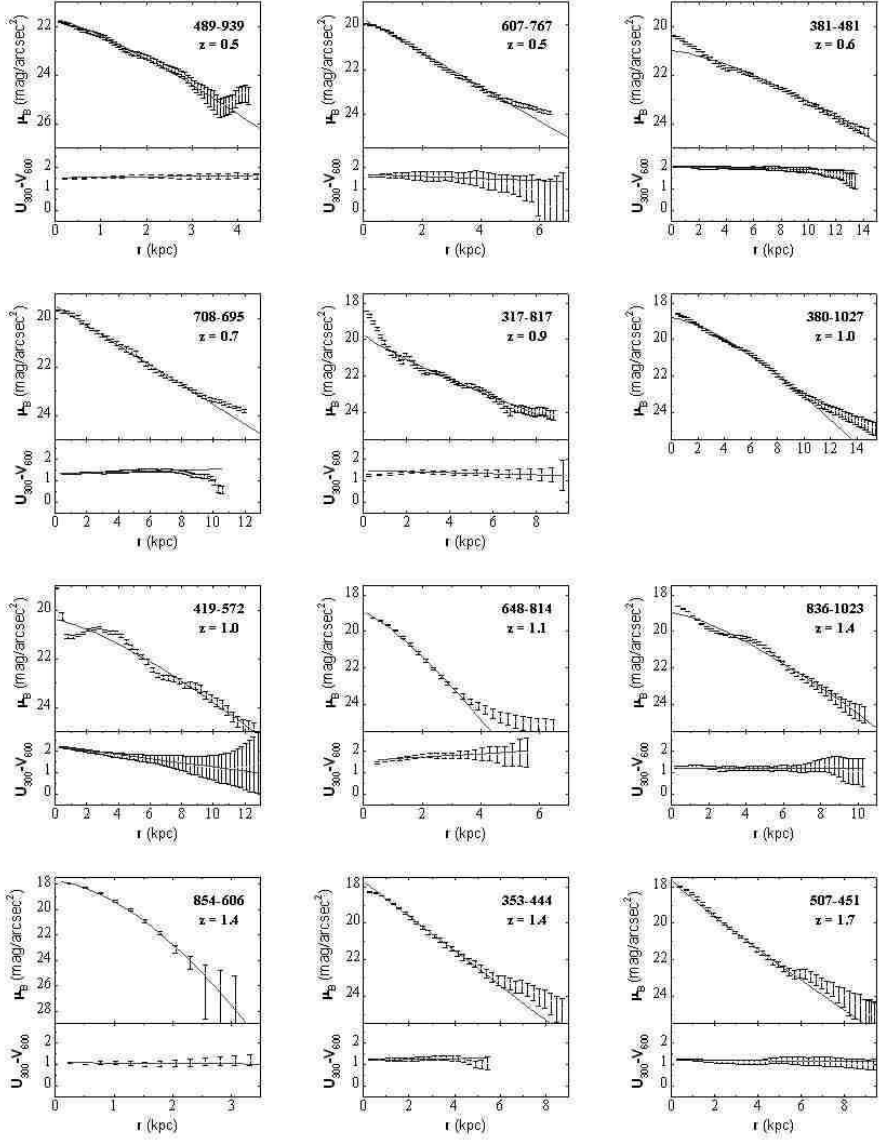}
\caption{Rest-frame $B$-band luminosity profiles with S\'{e}rsic
fits, and rest-frame $(U-V)$ color profiles with linear fits.}
\label{fig2}
\end{center}
\end{figure*}

\begin{figure*}
\begin{center}
\includegraphics[width=17cm]{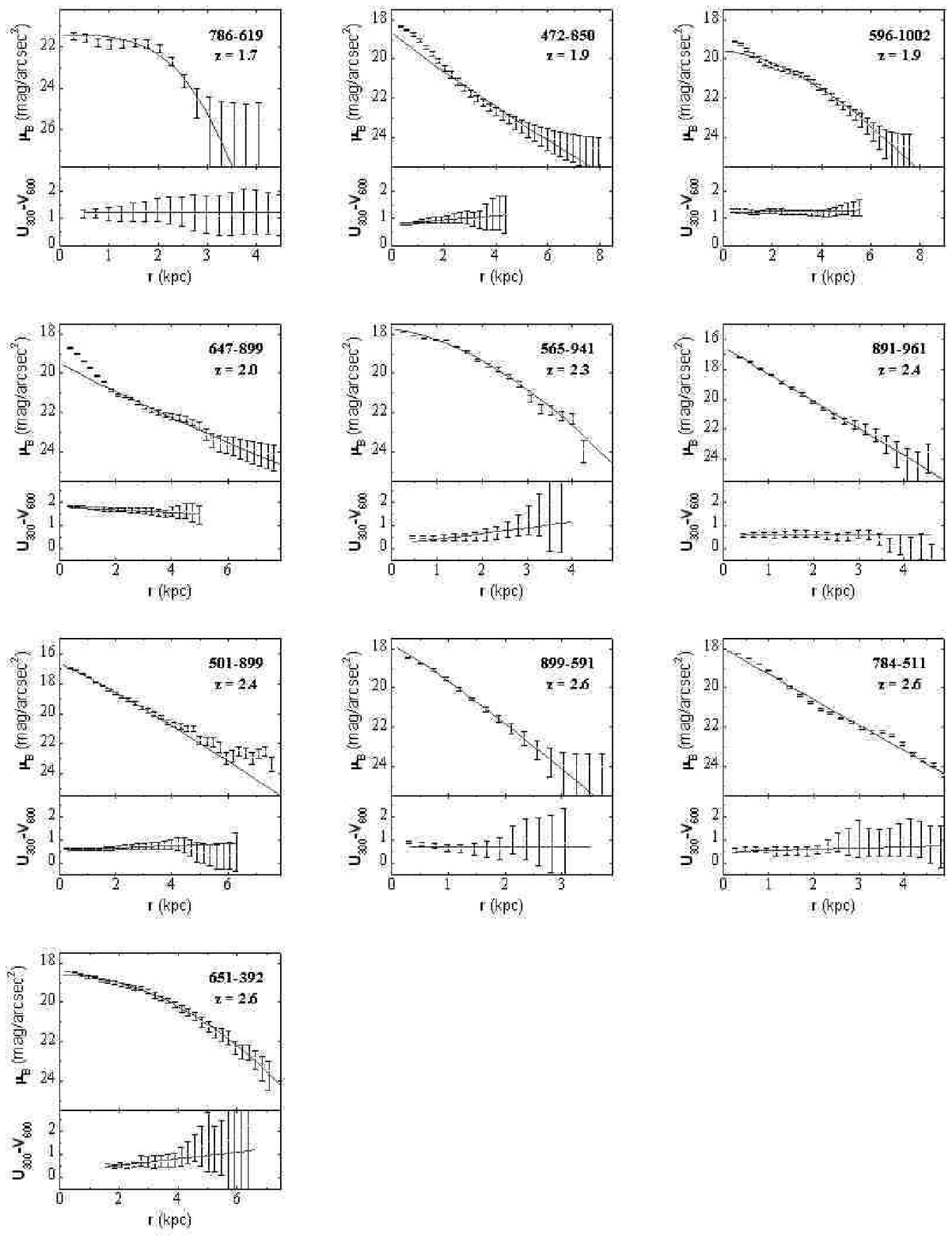}\\
\end{center}
Fig. 2. (Continued)
\end{figure*}

\section{Observations and sample selection}

To avoid generation of artificial evolutionary trends, galaxies at
different redshifts should be observed in the same rest-frame
waveband. This precaution eliminates possible effects of
``morphological k-correction", i.e. the dependence of the
photometric properties of a galaxy on the rest-frame waveband. This
phenomenon is especially frequent in the case of mid-type galaxies
(Papovich et al. \cite{papo}). For comparison with near-by galaxies,
rest frame optical passband would be the best choice. To study
rest-frame optical properties of galaxies at high redshifts,
observations made in near infrared are required. Presently, the best
available resolution in near infrared is offered by the NICMOS
camera aboard the Hubble Space Telescope (HST). However, the imaging
properties of the NICMOS camera are still far behind those of the
HST optical cameras.

During HDF South observations (Williams et al. \cite{will}), the
NICMOS camera was pointing in a slightly different direction, and in
this way, a parallel field was created with NIC-3 camera through the
$J,$ $H$ and $K$ broad-band infrared filters. This field is also
covered by WFPC2 camera $I$ filter observations of Flanking Field 9
(Lucas et al. \cite{luca}). Additionally, the NICMOS parallel field
has been observed with the HST STIS camera using the open mode, i.e.
without any filter, which gives a very broad bandwidth with the
central wavelength matching that of the standard $V$-band.
Regrettably, the $I$ and $K$ band observations could not be used
here -- the former suffers from insufficient depth and the latter
has too high noise and background level (the $K$ images were taken
during "bright" time, i.e. when the telescope was pointed near the
bright limb of the earth).

Observations conducted in the three remaining passbands -- $V,$ $J$
and $H$ -- have been used in the present study. The basic properties
of these observations and the final images are given in
Table~\ref{tab1}.

We have used the fully calibrated, combined and dithered images of
the HDF-S NICMOS parallel field, available via the homepage of the
Space Telescope Science Institute. The lower limit of detectable
surface brightness on these exposures is approximately 27.5
mag/arcsec$^2$.

Table~\ref{tab1} shows that on NICMOS exposures the full-weight
half-maximum (FWHM) of the point-spread function (PSF) is 0.23$''$,
causing serious distortion of the galactic images. The width of the
PSF as well as its slightly non-circular shape cause uncertainties
in deconvolution of the images.

Unfortunately, spectroscopic redshifts of the galaxies in the HDF-S
NICMOS parallel field have not been measured. Photometric redshifts,
based on 9-band measurements, have been calculated by Yahata et al.
(\cite{yaha}).

Galaxies with redshift $z>$ 0.5 were selected for the further study.
Discrimination between Hubble types is not always a simple task at
high redshifts. Many objects may be in transition stages (protodisks
or protospheroids) even at $1 < z < 2$ (Conselice et al.
\cite{cons9}). Morphological structure of these objects was
inspected visually on the basis of $H$-band exposures, all galaxies
suspected for being disks were included in the sample. In some
cases, also the high-resolution STIS $V$-band images were checked
(see Table 1 for the comparison of PSFs and projected pixel sizes).
Final confirmation for the sample to consist of  disk galaxies comes
from the luminosity profiles, which all exhibit S\'{e}rsic index
$n<2$ and should thus have a late-type morphology (Andredakis et al.
\cite{andre}, Ravindranath et al. \cite{ravi}).

Galaxies showing significant asymmetry or irregular shape were
excluded, because fitting ellipses to their isophotes would give
rather scattered light profiles (light distribution is very
sensitive to the galactic center position) and their interpretation
would not be straightforward in the context of the present models
and subsequent analysis. Thus possible starburst and interacting
galaxies were rejected as photometrically incomparable to regular
disks.

These selection criteria finally set a redshift limit at $z=2.6$,
beyond which no disk galaxies could be distinguished with acceptable
confidence. A final sample of 22 galaxies qualified for further
analysis.

The choice of passbands described above allows us to determine
rest-frame optical luminosity profiles, using STIS observations for
redshifts $z<1.0$ and NICMOS $J$ and $H$ observations for
$1.1<z<2.0$ and $z>2.1$, respectively; the mean central rest-frame
wavelength thereby becomes 420 nm, corresponding to Johnson $B$
filter. Color information can be obtained, using STIS and NICMOS $J$
observations at $z<1.1$, and NICMOS $J$ and $H$ bands at $z>1.4$ (no
galaxies were found at $1.1<z<1.4$). The mean central rest-frame
wavelengths thus become 350 nm and 580 nm, allowing the derivation
of $(U-V)$ color distribution. Note that here STIS observations are
used for $U$ and $B$ rest-waveband photometry for the same galaxies.
This can be justified by the very wide ``passband" of STIS open
mode, depending only on the detector sensitivity. A drawback of
using such a wide wavelength range lies in the danger of suppressing
possible color-features and trends, but it does not introduce or
artificially amplify the evolutionary effects.

The STIS and NICMOS $J$-band images of the galaxies are presented
in Fig~\ref{fig1}. STIS images are designated as $V$; $x$ and $y$
pixel coordinates of the galaxies on NICMOS images are used as
galactic names; $z$ is photometric redshift. Where necessary, the
galaxies of the present sample have been encircled to avoid
confusion. Some of the sample galaxies lie at the edge of the
field of view of the STIS camera and the galaxy 380-1027 falls
just outside it, thus no color-information could be acquired in
the latter case.

\section{Data reduction and photometry}

To determine absolute magnitudes and especially to generate color
profiles, it is vital to precisely estimate the background level for
each galaxy. We measured background level at $2-4$ empty-looking
fields around each galaxy and found the uniformly subtracted
background levels of both STIS and NICMOS final images to be
slightly over-estimated. Individual background estimates for each
galaxy were necessary due to considerable variations within each
passband image, ranging from $-0.00015$ to $0$ counts per second in
the case of STIS and from $-0.00005$ to $0.00002$ counts per second
in the case of NICMOS images.

Before fitting ellipses to the galactic images with the STSDAS
task ELLIPSE, we deconvolved images with PSF. Correct
deconvolution is very important, if one wishes to conduct analysis
of the luminosity and color profiles of faint objects. The bulge
component of a disk galaxy often acts as a point source at high
redshifts; wrong estimating of the spreading of its flux may cause
significant drift of the parameters of the disk component.

In the case of the NICMOS images, the model PSFs were created
using stellar images found in the same field. In the case of STIS,
no unsaturated stellar images were found and the model PSF was
created using the TinyTim package.

Ellipses were interactively fitted to the isophotes. Any object,
which did not seem to be part of a given galaxy, was masked. Both
fixed and free wandering values were tested for ellipse centers;
the resulting differences in surface brightness profiles are
included in the uncertainties of the profiles in Fig.~\ref{fig2}.
Ellipticity and position angle were kept fixed according to the
outer isophotes of the galaxies.

For rest-frame photometry we transformed the observed passbands
into rest-frame standard passbands, which needed minimal
k-correction after redshifting. $B$-profiles and $(U-V)$ color
profiles were found to be most suitable (see Sect. 2).

No reliable k-corrections existed for the transformations from
neither the STIS open mode nor the NICMOS passbands and we had to
calculate them. In our calculations we relied on the synthetic
spectra of redshifted Sb galaxies (as a mid-way between S0 and Sc
galaxies) constructed by Bicker et al. (\cite{bick}) according to
their chemical evolution models. In these spectra, effects of
evolution and redshifting had already been taken into account
(therefore the correction actually includes also evolution
correction). Our share was to calculate the relation between the
observed flux and the redshifted standard $U$, $V$ or $B$ filter
flux for each set of cameras, filters and redshifts. The throughput
curves of the NICMOS filters and STIS clear imaging were taken from
NICMOS and STIS Instrument Handbooks, respectively. Due to the
available choice of filters, the final k-corrections for both
rest-frame $B$-band and $(U-V)$ color remained rather modest,
typically around $0.4-0.7$ magnitudes.

The final, k-corrected surface brightness profiles in rest-frame
$B$-color with estimated error bars are presented in the upper
panels of Fig~\ref{fig2}. In the lower panels, k-corrected
rest-frame $(U-V)$ profiles and corresponding error bars are given.
To calculate the extent of the error bars, inaccuracy of our Hubble
type classification and the population synthesis models were taken
into account. The latter was difficult to estimate precisely because
no actual uncertainties of chemical evolution models are available.
Thus we made a rough estimate on the basis of deviations between
synthetic spectra calculated by different authors. These
uncertainties are amplified by different widths and shapes of the
filter passbands. The uncertainties of the synthetic spectra at
higher redshifts are larger than at lower redshifts, but the
estimated errors remain comparable ($0.25-0.45$ mag). This is
because of the spectra being stretched as $(1+z)$ with respect to
the filter passband on one hand, thus enabling more exact
determination of the flux, and the usage of the wide STIS imaging
throughput at the tricky $UV$-region for lower redshift galaxies on
the other hand, which can not be very accurate. The step of
determining actual flux according to the synthetic spectra dominates
in the estimates of the uncertainties of color measurements on Fig
~\ref{fig8}.

The usage of different filters and detectors for imaging at
different redshifts would cause errors in absolute photometry and
introduce artificial trends with respect to redshift, if
calibrations or k-corrections were calculated improperly.
Considerable errors of this kind would be seen as a jump in
magnitude-redshift plots at the redshift of the filter change,
e.g. between $z=1.1$ and $z=1.4$ in our case of $(U-V)$ vs. $z$
graph in Fig~\ref{fig6}. The graph of absolute $B$ magnitudes
$M_B$ vs. $z$ was also checked for this effect; no jumps were
found.

\section{Fitting by model profiles}

Although in some cases (usually at smaller redshifts), the bulge
component was distinguishable from the disk, no attempt was made
to split the luminosity profile into two components, because this
would have given systematically different disk parameters at
different redshifts.

Luminosity profiles were fitted by a S\'{e}rsic surface density
distribution (S\'{e}rsic, \cite{sersic}):
\begin{equation}
I(a) = I(0) \exp [ -b_n (r/r_e)^{1/n}], \label{ser}
\end{equation}
In this formula, $r$ is the distance along the galactic major
axis, $r_e$ is the effective radius, containing half of the total
luminosity, $n$ is the shape parameter, setting the curvature of
the profile, and $b_n$ is a normalizing constant, dependent on
$n$, enabling to keep $r_e$ the half-light radius. To determine
$b_n$, either gamma functions or integral equations have to be
numerically solved. Usually, $b_n = 1.9992n-0.3271$ (Capaccioli
\cite{capa}) or other formulae are used (e.g. Prugniel \& Simien
\cite{pr:si}, Moriondo et al. \cite{mori}, Ciotti \& Bertin
\cite{ci:be}), which all can be successfully applied at $n$ values
above $n\approx 0.5$. If lower $n$ values are present, these
solutions start to mislead. Therefore, we have calculated our own
approximation,
\begin{equation}
b_n = 2n - {1\over 3} + {1\over 65n}, \label{bn}
\end{equation}
which can well be used down to $n\approx0.1$. Comparison of our
approximation (\ref{bn}) with numerically calculated values together
with approximations by Capaccioli \cite{capa} and Ciotti \& Bertin
\cite{ci:be} is given in Fig.~\ref{fig4}. To emphasize deviations at
small $n$ values, logarithmic scale is used.

\begin{figure}
\resizebox{\hsize}{!}{\includegraphics{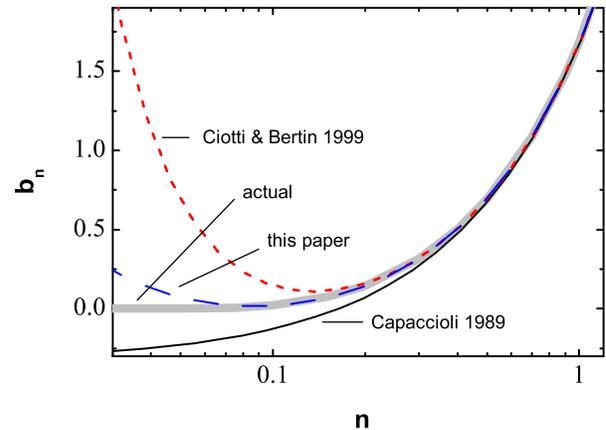}}
\caption{Normalizing parameter $b_n$ in S\'{e}rsic formula
(\ref{ser}) at low $n$ values. Comparison of our approximation
(\ref{bn}) with numerically calculated  actual values, together with
approximations by Capaccioli \cite{capa} and Ciotti \& Bertin
\cite{ci:be}.} \label{fig4}
\end{figure}

As a tool for fitting surface brightness distribution, the
S\'{e}rsic law has a purely empiric background. Starting from a
spatial density distribution law and projecting it along the line of
sight would be a more physical method, also allowing a
straightforward comparison with kinematic data and thereby enabling
the construction of self-consistent models for mass distribution.
Thus, fitting of a surface density distribution, deduced from the
space density distribution law of the very general form
\begin{equation}
\rho (a)=\rho (0)\exp [ -( a/(ka_0))^{1/n} ] , \label{ein}
\end{equation}
was also tested for comparison. Here, $\rho (0)$ is the central
density, $a= \sqrt{R^2+z^2/q^2}$, where $R$ and $z$ are two
cylindrical coordinates and $q$ is the axial ratio; $a_0$ is the
characteristic radius and $k$ is a normalizing parameter, similar to
S\'{e}rsic $b_n$ (see Einasto \& Haud \cite{ei:ha} and Tenjes et al.
\cite{tenj} for more details about this distribution and reasoning
for its usage). This method gives the luminosity profile a somewhat
different shape, but if fitted to a real density distribution, the
main parameters ($n$, $a_0\approx r_e$) remain similar to those
acquired from the S\'{e}rsic law.

Comparison of these two distributions at different $n$ values is
given in Fig~\ref{fig5}. Surface brightness value $\mu$ and radius
$r$ are given in the same relative units for both distributions.
We stress that no rescaling has been conducted for matching the
distributions; the fit has been achieved by keeping the total
luminosities equal. Note that a perfectly exponential
distribution, corresponding to the S\'{e}rsic law with $n$=1
cannot be attained with distribution (\ref{ein}).

For galaxies in the present sample, no additional kinematic data are
available at present. For this reason, we only present here the
parameters resulting from fitting of the S\'{e}rsic law (\ref{ser})
to the brightness profiles. This also allows more direct comparison
with other studies.

\section{Results and discussion}

\begin{table*}
   \caption[]{General galactic parameters.}
   \label{tab2}
\begin{center}
\begin{tabular}{lllllllll}
\hline
Name$^a$    &  NED name            & $z^b$ & $M_B$     & $L_B$ & $r_e$     &  $n^c$&$U-V$      & $\Delta(U-V)/\Delta r$  \\
        &  (also J2000 coordinate) & & $\rm(mag)$& $\rm 10^{10}L_{\sun}$ & $\rm (kpc)$&&$\rm (mag)$& $\rm (mag/kpc)$   \\
\hline
489 -   939 &   HDFS J223252.31-603827.3    &   0.49    &   -16.4   &   4.3 &   1.8 &   0.8 &    1.55   &   0.03    \\
607 -   767 &   HDFS J223251.11-603840.2    &   0.51    &   -17.7   &   3.3 &   2.3 &   1   &    1.40   &   -0.05   \\
381 -   481 &   HDFS J223253.42-603901.6    &   0.6     &   -18.8   &   3.2 &   6.7 &   0.7 &    1.66   &   -0.01   \\
708 -   695 &   HDFS J223250.08-603845.6    &   0.66    &   -19.3   &   3.2 &   4.5 &   1   &    1.26   &   0.02    \\
317 -   817 &   HDFS J223253.65-603846.9    &   0.89    &   -18.3   &   1.9 &   3.0 &   1.2 &    1.28   &   -0.01   \\
380 -   1027&   HDFS J223253.43-603820.6    &   0.99    &   -20.8   &   1.8 &   4.1 &   0.7 &    ~~~~ $-$& ~~~~ $-$ \\
419 -   572 &   HDFS J223253.03-603854.8    &   1.01    &   -20.1   &   1.7 &   4.8 &   0.75&    1.69   &   -0.10   \\
648 -   814 &   HDFS J223250.70-603836.6    &   1.12    &   -18.6   &   1.6 &   1.3 &   0.8 &    1.66   &   0.08    \\
836 -   1023&   HDFS J223248.78-603821.0    &   1.37    &   -21.1   &   1.3 &   3.5 &   0.75&    1.18   &   -0.01   \\
854 -   606 &   HDFS J223248.59-603852.2    &   1.37    &   -19.6   &   1.2 &   0.7 &   0.6 &    1.05   &   -0.02   \\
353 -   444 &   HDFS J223253.71-603904.4    &   1.39    &   -20.1   &   1.2 &   1.9 &   1   &    1.22   &   0.02    \\
507 -   451 &   HDFS J223252.13-603903.9    &   1.65    &   -20.2   &   1.0 &   2.0 &   1.1 &    1.14   &   -0.01   \\
786 -   619 &   HDFS J223249.29-603851.2    &   1.65    &   -17.2   &   0.8 &   1.4 &   0.3 &    1.41   &   0.00    \\
472 -   850 &   HDFS J223252.49-603833.9    &   1.89    &   -19.3   &   0.8 &   2.0 &   1.1 &    0.95   &   0.09    \\
596 -   1002&   HDFS J223251.22-603822.6    &   1.93    &   -20.0   &   0.8 &   2.6 &   0.6 &    1.23   &   0.01    \\
647 -   899 &   HDFS J223250.70-603830.3    &   2.01    &   -19.7   &   0.8 &   2.8 &   1.1 &    1.71   &   -0.06   \\
565 -   941 &   HDFS J223251.54-603827.1    &   2.33    &   -19.7   &   0.5 &   1.6 &   0.6 &    0.64   &   0.25    \\
501 -   899 &   HDFS J223252.19-603830.2    &   2.39    &   -20.8   &   0.4 &   1.7 &   0.9 &    0.64   &   0.05    \\
891 -   961 &   HDFS J223248.22-603825.6    &   2.39    &   -20.8   &   0.3 &   1.0 &   1   &    0.61   &   0.00    \\
899 -   591 &   HDFS J223248.13-603853.4    &   2.55    &   -19.3   &   0.2 &   0.9 &   0.9 &    0.62   &   0.00    \\
784 -   511 &   HDFS J223249.30-603859.4    &   2.56    &   -19.2   &   0.1 &   1.5 &   1   &    0.58   &   0.05    \\
651 -   392 &   HDFS J223250.67-603908.3    &   2.58    &   -19.8   &   0.1 &   2.7 &   0.5 &    0.59   &   0.20    \\

 \hline
\end{tabular}
\end{center}
\begin{list}{}{}
\item[$^a$] The $x$ and $y$ pixel coordinate on HDF-S NICMOS final
images
\item[$^b$] Photometric redshifts are from Yahata et al.(\cite{yaha}).
\item[$^c$] Steepness parameter of the best-fit S\'{e}rsic profile.
\end{list}
\end{table*}

The rest-frame $B$ luminosity profiles, fitted to the S\'{e}rsic
law, and the $(U-V)$ color profiles with linear fits are shown in
Fig~\ref{fig2}. The main properties of the galaxies and the results
of the photometric analysis are given in Table~\ref{tab2}.
Distribution of the parameter $n,$ the effective radius $r_e$, the
$(U-V)$ color distribution, and color gradient $\Delta(U-V)/\Delta
r$ are shown in Figs.~\ref{fig6} -- \ref{fig8} as a function of
redshift. The three different brightness levels of the data points
indicate three absolute rest-frame $B$ magnitude-bins of the
galaxies: white points stand for galaxies with $M_B < -20$ mag, grey
points for $M_B=-19...-20$ mag and black points for $M_B > -19$ mag.
For equal treatment of the galaxies at different redshifts, absolute
brightness has been calculated within rest-frame surface brightness
$\mu_B<24$ mag/arcsec$^2$.

\begin{figure}
\resizebox{\hsize}{!}{\includegraphics{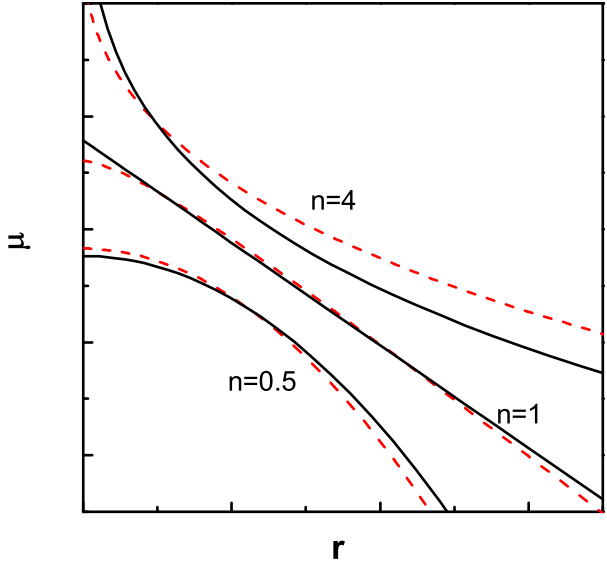}}
\caption{Comparison of the S\'{e}rsic surface brightness
distribution (solid lines) and a distribution, derived from the
space density distribution of Eq.~\ref{ein} (dashed lines) at
different $n$ values. Surface brightness value $\mu$ and radius
$r$ are given in the same relative units for both distributions.
No rescaling has been conducted to match the curves.} \label{fig5}
\end{figure}

\begin{figure}
\resizebox{\hsize}{!}{\includegraphics{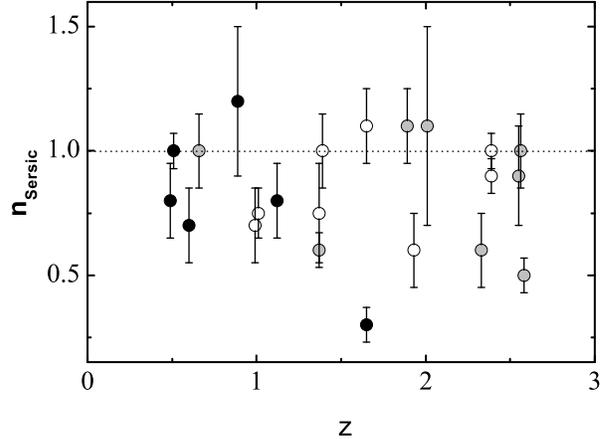}} \caption{The
S\'{e}rsic luminosity profile curvature $n$ as a function of
redshift. Open circles stand for galaxies with absolute luminosity
$M_B < -20$ mag, gray circles for $-19$ mag $< M_B < -20$ mag and
black circles for $M_B < -19$ mag.} \label{fig6}
\end{figure}

\begin{figure}
\resizebox{\hsize}{!}{\includegraphics{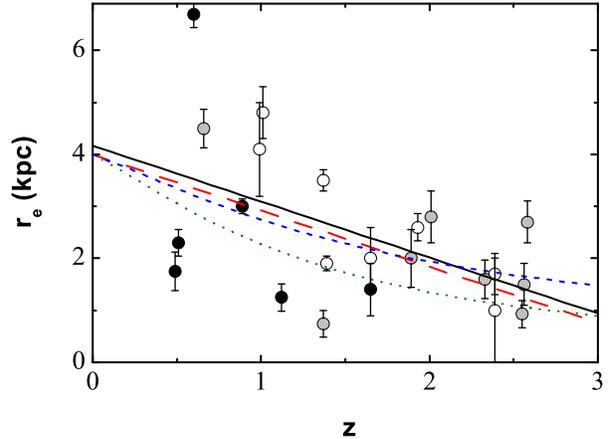}} \caption{The
S\'{e}rsic luminosity profile effective radius $r_e$ as a function
of redshift. Open circles stand for galaxies with absolute
luminosity $M_B < -20$ mag, gray circles for $-19$ mag $< M_B <
-20$ mag and black circles for $M_B < -19$ mag. The solid line
shows a least-squares linear fit to the current sample
$r_e(z)/r_e(0) = 1-0.26z$; the long-dashed line is the relation
$r(z)/r(0) = 1-0.27z$ derived by Bouwens \& Silk (\cite{bo:si});
the dotted and short-dashed lines give the relations $r_e\sim
H(z)^{-1}$ and $r_e \sim H^{-2/3}(z)$, respectively.} \label{fig7}
\end{figure}

\begin{figure}
\resizebox{\hsize}{!}{\includegraphics{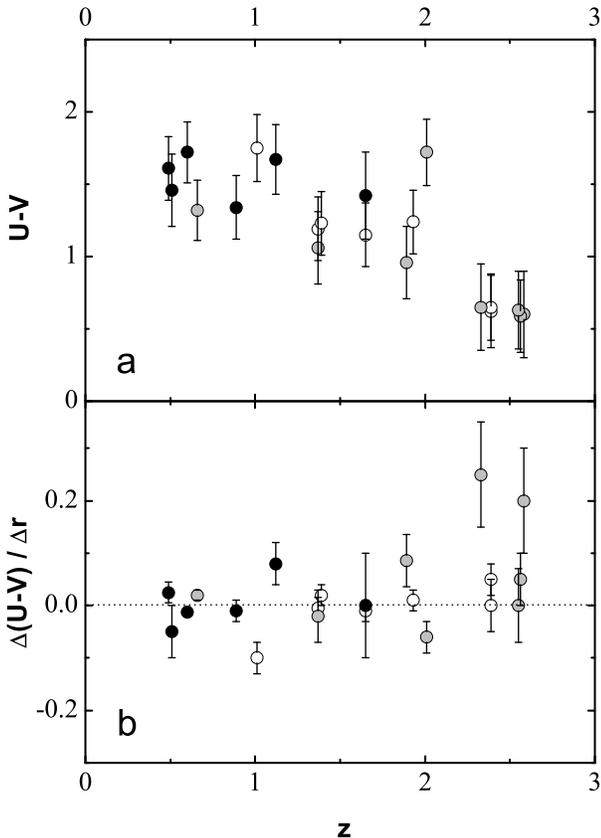}} \caption{Galactic
parameters -- rest frame $(U-V)$ color (a) and $\Delta(U-V)/\Delta
r$ (b) -- as a function of redshift. Designations are the same as in
Fig.~\ref{fig7}}\label{fig8}
\end{figure}

\begin{figure}
\resizebox{\hsize}{!}{\includegraphics{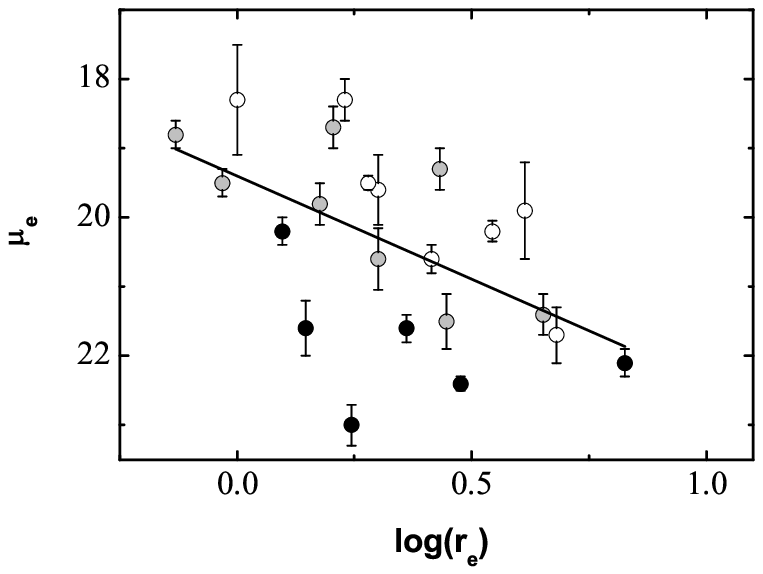}}
\caption{Effective surface brightness $\mu_e$, plotted as a
function of $log(r_e)$. Designations are the same as in
Figs.~\ref{fig6}, \ref{fig7}. Solid line shows a least-square
fit.}\label{fig9}
\end{figure}

\subsection{The shape of the luminosity profiles}

As is evident from the present sample, nearly half of galaxies
cannot be fit to an exponential disk model, while for the rest,
$n\approx$ 1 works well out to the highest redshifts
(Fig.~\ref{fig6}). A similar result was obtained by Moth \& Elston
(\cite{moth}) for the HDF North field. As suggested in Tamm \&
Tenjes (\cite{TT2}), the more curved profiles might indicate the
effect known as disk truncation, discovered by studies of local
edge-on disk galaxies (de Grijs et al. \cite{grijs}, Pohlen et al.
\cite{pohl}) and more recently detected also at higher redshifts
(P\'{e}rez \cite{pere}, Trujillo \& Pohlen \cite{tr:po}). Such a
luminosity distribution is sometimes fitted by a double-exponential
profile (Pohlen et al. \cite{pohl}, P\'{e}rez \cite{pere}). The
physical background of disk truncation has not become clear as yet
(see e.g. Sasaki \cite{sasa} and de Grijs et al. \cite{grijs} for
more discussion about the origin of this phenomenon).

In contrast to these truncated disks, some galaxies in the present
sample have flattening profiles (e.g. $607-767$, $708-695$,
$380-1027$, $353-444$). Similar profiles have been measured near
$z=$1 for gamma-ray burst-selected disks (Conselice et al.
\cite{con2}) and out to $z=$3 for large disks in HDF-South primary
field (Labb\'{e} et al. \cite {labb}). Erwin et al. (\cite{erwin})
have found a significant proportion of local barred S0-Sb galaxies
with flattening profiles, which they call ``anti-truncated" disks.
They claim that, at least in several cases, this effect might be
caused by interactions. Observations of interacting galaxies (Chitre
\& Jog \cite{ch:jo}) and numeric simulations (Bournaud et al.
\cite{bour}) have demonstrated the possibility of mergers to smear
out the outer parts of luminosity profiles. In the case of our
sample, visual companions are present for all four galaxies with
flattened profiles, thus interactions may have a role in the
development of this flattening. The companions have been masked
prior to ellipse fitting, therefore the luminosity of the companions
cannot be responsible for the effect.

The 7 galaxies studied in Tamm \& Tenjes (\cite{TT1}, \cite{TT2})
have significantly low $n$ values compared to the present sample.
This is most likely a result of the small size of the samples or
selection effects, favoring luminous galaxies with regular
kinematics.

\subsection{Sizes of disks $r_e$}

In Fig.~\ref{fig7}, evolution of the size of the present sample is
shown. Error bars of sizes were derived by varying $r_e$ while
fitting the S\'ersic law within the surface brightness profile error
bars. The slight degeneracy between parameters $n$ and $r_e$ has
also been taken into account.

Starting from the locally observed Universe and modeling galactic
evolution backwards, Bouwens \& Silk (\cite{bo:si}) derived a
scaling relation $r(z)(r(0) = 1-0.27z$ for the $B$-band radius. The
long-dashed line in Fig. \ref{fig7} shows this relation. The
least-squares linear fit to our sample gives $r_e(z)/r_e(0)=1-0.26z$
(solid line). Here, the median value of S\'{e}rsic half-light radii
$\langle r_e\rangle\approx4.2$ kpc for late-type galaxies from the
SDSS survey (Shen et al. \cite{shen}) has been used as the value of
$r_e(0)$ (considering the median absolute luminosity of the present
sample). It is seen that the model developed by Bouwens \& Silk
(\cite{bo:si}) fits our data rather closely. Our result is
consistent with the small $r_e$ values found for UDF galaxies by
Elmegreen et al. (\cite{elme}).

Galactic size evolution has also been predicted from simple models
of the hierarchical structure formation scenario and an approximate
scaling has been derived with disk formation time: $r_e\sim V_{\rm
vir}/H(z)$, where in the case of $\Omega_0=1$, $H(z) = H_0 [\Omega_m
(1+z)^3 + \Omega_{\Lambda} ]^{1/2}$ (Mo et al. \cite{mo}). In a
study of galaxies at a redshift range $z = 1.5 - 5$, Ferguson et al.
(\cite{ferg}) found the scaling of $r_e\sim H^{-1}(z)$ to give a
good fit to their sample, while Trujillo et al. (\cite{truj2})
achieved a better fit with $r_e \sim H^{-2/3}(z)$ for galaxies at $z
= 0.3-2.5$. The relations $r_e\sim H^{-1}(z)$ and $r_e \sim
H^{-2/3}(z)$, scaled with $r_e(0)\approx4$ kpc, are also shown in
Fig. \ref{fig7}, as dotted and short-dashed lines, respectively. The
former shows a decrease significantly faster than that of the
present sample. However, it must be kept in mind that the
theoretical relation uses the disk formation redshift instead of the
observed redshift for scaling. Further more, in contrast to the
present sample and the literature referred to above, Ravindranath et
al (\cite{ravi}) and Cassata et al (\cite{cass}) did not detect any
significant evolution of galaxy sizes with redshift in a thorough
study of galaxies at $z \le 1.2$.

Fig. \ref{fig7} shows that the trendline fitted to the present
sample is quite sensitive to the data point corresponding to the
largest galaxy with $r_e =$ 6.7~kpc at $z=0.6$. Its extensive disk
is obvious also on Fig. \ref{fig1} and its light profile is a very
regular one, thus there is no doubt that this galaxy is clearly a
large disk galaxy. The only reason for this data point to be
significantly offset could be a wrong photometric redshift estimate.

\subsection{Color $(U-V)$ and color gradient $\Delta(U-V)/\Delta r$}

In Fig.~\ref{fig8}a, rest frame $(U-V)$ color as a function of
redshift is presented. The $(U-V)$ color shows a mild decrease until
$z\approx2$, followed by a notable drop of roughly 0.5 mag by
$z\approx2.5$. At lower redshifts, the dependence can be compared to
the Deep Groth sample studied by Weiner at al. (\cite{wein}), for
which a very slight, if any, decrease of $(U-B)$ was detected for
late-type galaxies in the redshift range $z=0-1.5$. The notable drop
of the $(U-V)$ values at $z\approx2$ cannot be due to miscalibration
-- a uniform set of filters and calibrations has been used for the
redshift range $z=1.4-2.6$ (see Sects 2 and 5.5). The drop can
rather be related to a major star formation peak at $z>2$ (see
below). A decrease of the rest-frame $(U-V)$ color with increasing
redshift up to $z\sim 3$ was derived also by Kajisawa \& Yamada
(\cite{ka:ya}). More detailed comparison with our results is not
possible at present because in their study, different morphological
types are presented together. Our result is quantitatively close
also to the result obtained from the analysis of the synthetic
spectra of Bicker et al. (\cite{bick}). Absolute $(U-V)$ values are
rather sensitive to k-correction, which, in turn, is uncertain. This
may cause a constant vertical shift of $(U-V)$ values. However,
trends in Fig.~\ref{fig8} should be independent of these
uncertainties.

Fig.~\ref{fig8}b shows that at redshifts $z \le 2$, there are small
or no color gradients. Gradients begin at $z>2$ and are dominantly
positive. A similar rise with redshift of the rest-frame color
$(U_{218}-U_{300})$ gradients was discovered by Moth \& Elston
(\cite{moth}), suggesting that star-formation was more centrally
concentrated at higher redshifts.

Recent hydrodynamical simulations of galaxies in $\Lambda$CDM
cosmology by Robertson et al. (\cite{robe}) indicate that star
formation in disks peaks between redshifts $z = 2-4$. Studies of the
Fundamental Plane of early-type galaxies at redshifts $z \le 1.3$
have also shown that the last epoch of major star formation peaks at
$z\sim 2-3.5$ (see Holden et al. \cite{hold} and references
therein). A jump in star formation beyond redshift $z = 2$ matches
well with the jump in $(U-V)$ color in Fig.~\ref{fig8}a (see above).

Smaller color gradients at lower redshifts $z = 0.5-2$ are due to
smoothing by interactions in disk evolution (Conselice et al.
\cite{cons}) and large scale gas motions (Samland \& Gerhard,
\cite{sa:ge}).

Positive color gradients also exist in local disk galaxies
(MacArthur et al. \cite{maca}). Positive color gradients may
indicate disk formation from inside out. For the Milky Way
protogalaxy such formation scenario was suggested by van den Bergh
(\cite{berg}). In the case of high-redshift galaxies, direct
interpretation of color gradients may be rather complicated. Color
gradients are determined by radial distribution of initial
metallicities, stellar ages, star-formation rate, gas accretion
details etc. Dependently on model details, both outside-in and
inside-out formation of disks can be simulated (Sommer-Larsen et
al. \cite{som-lar}; Robertson et al. \cite{robe}).

According to the analysis of their disk galaxy formation model,
Westera et al. (\cite{west}) have found that luminosity from $U$
to $V$ mainly indicates star formation, and the effects of age and
metallicity are negligible. Thus the positive color gradient most
possibly refers to intensive star formation in central parts of
galaxies at $z>2$.

Relying on $(U-V)$, we may roughly estimate the disk formation
time. According to our photometry, the disks at $z\sim 2.5$ have
$(U-V)\sim 0.7.$ On the basis of simple stellar population
chemical evolution models (Worthey \cite{worthey}; Bressan et al.
\cite{bres}), these disks have ages $1-2$ Gyr and have thus formed
at $z=3.5-7$.

\subsection{Surface brightness at effective radius}

Coenda et al. (\cite{coen}) have shown that disks obey similar
photometric scaling relations as spheroids and elliptical galaxies.
Degeneracy exists between the effective radius, effective surface
brightness, the parameter $n$ and absolute magnitude. For the
current sample, a trend can be shown for the effective surface
brightness $\mu_e$ to diminish with larger effective radius $r_e$,
presented in Fig.~\ref{fig9}. Least squares fit gives $\mu_e =$
$3.0\log r_e + 19.4$ mag, which, regarding the small size of the
sample, is remarkably close to the dependence found by Coenda et al.
(\cite{coen}) $\mu_e =$ $3.4 \log r_e + 18.7$ mag .

\subsection{Uncertainties of the study}

Small sample sizes and selectional biases are critical topics for
most of the studies of high-redshift galaxies. The present sample of
22 galaxies can not offer reliable statistics and cautions against
jumping into severe conclusions.

Morphological identification of these disk galaxies was obtained by
visual inspection of $H$-band exposures, all galaxies suspected for
being disks were included in the sample. In some cases, also the
high-resolution STIS $V$-band images were checked. Some of the
galaxies of the present sample have been studied on the basis of
NICMOS images by Rodighiero et al. (\cite{rodi}), and classified as
E-S0. We would like to point out that for classification and
accurate photometry it is important to carefully deconvolve the PSF
of NICMOS images. Even in the case of good restoration of NICMOS
images, additional observations with other detectors, preferably
with higher resolution (e.g. STIS in the present case) still prove
beneficial. For example, visual inspection of the STIS image of the
galaxy $419-572$ reveals its nature of a late-type (at least Sb)
galaxy.

Usually, the cosmological dimming as $(1+z)^4$ favors the selection
of more luminous galaxies at higher redshifts. For our sample,
Figs~\ref{fig6}--\ref{fig9} reveal no clear dependence of any of the
parameters on the absolute luminosity of the galaxies; this fact
allows us to consider the sample to be sufficiently complete for the
conducted analysis.

The selection of the sample on the basis of $H$-band imaging
(corresponding to $B$-band at $z>2$) might cause a slight bias
towards bluer galaxies at higher redshifts. On the other hand, the
change of the spectral energy distribution is slow between $B$ and
$R$ passbands and we consider this possible selection effect not to
be a reason of a color jump at $z\simeq 2$ in Fig.~\ref{fig8}.

For calculating k-corrections, all galaxies were assumed to be Sb
morphological types. This approximation causes additional scatter of
absolute magnitudes and colors. However, according to our estimates,
the resulting uncertainties remain within 0.2 magnitudes even in the
worst cases.

Among possible sources of errors and uncertainties, the ever
insufficient imaging depth and spatial resolution should be
considered. With the currently available NIR imaging a dilemma
remains, whether to consider faint patches around galaxies to belong
to the galaxy and include them in photometry or to mask them as
extraneous. This may slightly affect the luminosity distribution at
the outer regions of a given galaxy and the values of the
photometric parameters, especially the shape parameter $n$ . In the
present case, we have followed a rather conservative masking
strategy, excluding most of the ``suspicious" patches. A more
liberal treatment would give slightly higher $n$ and $r_e$ values
and lower absolute magnitudes. The shift of these values would be
rather systematic.

Insufficient spatial resolution influences the interpretation of
surface brightness profiles in the central parts of galaxies. In the
present study we decided not to decompose galaxies into the bulge
and disk components. Otherwise, rather serious systematic
differences would appear in handling of low-redshift and
high-redshift galaxies. In general, decomposing light profiles into
a share of a bulge and a disk would decrease the disk radius.

\begin{acknowledgements}

We would like to thank the anonymous referee for useful comments and
suggestions helping to improve the paper. We acknowledge the
financial support from the Estonian Science Foundation (research
grant 6106). This paper is based on NASA/ESA Hubble Space Telescope
NICMOS and STIS observations of the Hubble Deep Field South obtained
from the data archive at the Space Telescope Science Institute. This
research had made use of the NASA/IPAC extragalactic database (NED),
which is operated by the Jet Propulsion Laboratory, California
Institute of Technology, under contract with the NASA.

\end{acknowledgements}

\end{document}